\documentclass{article}
\usepackage{times}
\usepackage{amsfonts}
\usepackage{graphicx}
\usepackage[pdfmark]{hyperref}
\begin{document}
\noindent
{\Large  A MECHANICS FOR THE RICCI FLOW}
\vskip1cm
\noindent
{\bf S. Abraham}${}^{1}$, {\bf P. Fern\'andez de C\'ordoba}${}^{2}$, {\bf Jos\'e M. Isidro}${}^{3}$ and {\bf J.L.G. Santander}${}^{4}$\\
Grupo de Modelizaci\'on Interdisciplinar, Instituto de Matem\'atica Pura y Aplicada,\\ Universidad Polit\'ecnica de Valencia, Valencia 46022, Spain\\
${}^1${\tt sabraham@mat.upv.es}, ${}^2${\tt pfernandez@mat.upv.es}, \\
${}^3${\tt joissan@mat.upv.es}, ${}^4${\tt jlgonzalez@mat.upv.es}
\vskip.5cm
\noindent
{\bf Abstract} We construct the classical mechanics associated with a conformally flat Riemannian metric on a compact, $n$--dimensional manifold without boundary. The corresponding gradient Ricci flow equation turns out to equal the time--dependent Hamilton--Jacobi equation of the mechanics so defined.

\section{Introduction}\label{intt}

The Ricci flow has provided many far--reaching insights into long--standing problems in geometry and topology \cite{PERELMAN, TOPPING}. Perhaps more surprising is the fact that it also has interesting applications in physics, where two important occurrences are conformal gravity \cite{WEYL}, and the low--energy effective action of string theory \cite{DHOKER}. Recent works \cite{CARROLL1, CARROLL5, CARROLL2, CARROLL6} have shed light on applications of the Ricci flow to foundational issues in quantum mechanics. The present contribution is a continuation of previous research \cite{SERGIO, ISIDRO, RICCI, DICE} on the exciting interrelation between Ricci flow, quantum mechanics and gravity. We can state our results more precisely as follows:\\

\noindent
{\bf Theorem.} {\it Let\/ $\mathbb{M}$ be a smooth, $n$--dimensional, compact manifold without boundary, endowed with a conformally flat Riemannian metric at some initial time $t_0$ (time is not a coordinate on $\mathbb{M}$, but an external parameter). Then a classical mechanical system having $\mathbb{M}$ as its configuration space can be defined, such that its time--independent mechanical action (Hamilton's principal function) equals the conformal factor of the initial metric. Moreover, the corresponding gradient Ricci flow equals the time--dependent Hamilton--Jacobi equation of the mechanical system thus defined.}\\

\noindent
{\bf Corollary.} {\it On the manifold $\mathbb{M}$,  the action functional for Einstein--Hilbert gravity equals the sum of the action functional for Schroedinger quantum mechanics, plus Perelman's functional, plus the Coulomb functional (see eqns. (\ref{irreplaceble}), (\ref{culo})  below).}\\

\noindent
Proofs of the theorem and of the corollary will be given in sections \ref{buch} and \ref{diskk}, respectively. As we will see, a key role will be played by the conformal metric we place on the manifold at the initial time $t_0$, when the Ricci flow starts to run. Now, in $n=2$ dimensions any metric is conformal, so the requirement of conformality is void in this case. Not so, however, when $n>2$, where a metric need not be conformal. However, when the metric {\it is}\/ conformal, it is univocally determined by the knowledge of just one function, called  {\it conformal factor}\/, on the manifold $\mathbb{M}$. We will see that  the conformal factor plays a central role in identifying the sought--for mechanical model. In principle, the result summarised by the above theorem holds for a {\it classical}\/ mechanical model. However, quantising the latter will also allow us to work with a {\it quantum}\/ mechanical model that is associated, in a canonical way,  with the initial conformal metric. \\

\noindent
{\bf Conventions.} Let a Riemannian metric $g_{ij}$ be given on $\mathbb{M}$. In local coordinates $x^{i}$, $i=1,\ldots, n$, the volume element  d$V$ on $\mathbb{M}$ equals
\begin{equation}
{\rm d}V=\sqrt{g}\,{\rm d}x^1\wedge\cdots\wedge{\rm d}x^n=:\sqrt{g}\,{\rm d}^nx,
\label{volele}
\end{equation}
where $g:=\vert\det g_{ij}\vert$. Using the Christoffel symbols of the corresponding Levi--Civita connection,
\begin{equation}
\Gamma_{ij}^m=\frac{1}{2}g^{mh}\left(\partial_ig_{jh}+\partial_jg_{hi}-\partial_hg_{ij}\right),
\label{toff}
\end{equation}
the Ricci tensor reads
\begin{equation}
R_{ik}=\partial_l\Gamma_{ik}^l-\partial_k\Gamma_{il}^l+\Gamma_{ik}^l\Gamma_{lm}^m-\Gamma_{il}^m\Gamma_{km}^l.
\label{riconventi}
\end{equation}
Given an arbitrary smooth function $\varphi$ on $\mathbb{M}$, the Laplacian $\nabla^2\varphi$ and the squared gradient $\left(\nabla\varphi\right)^2$ have the following coordinate expressions:
\begin{equation}
\nabla^2\varphi:=\frac{1}{\sqrt{g}}\partial_r\left(\sqrt{g}g^{rs}\partial_s\varphi\right),\qquad
\left(\nabla\varphi\right)^2:= g^{mn}\partial_m\varphi\partial_n\varphi.
\label{vdrei}
\end{equation}

In $n=2$ dimensions, any Riemannian metric is conformally flat; such was the case analysed in refs. \cite{SERGIO, RICCI}. In $n\geq 3$ dimensions, conformality of the metric is no longer guaranteed. Specifically \cite{GOLDBERG}, when $n=3$, a necessary and sufficient condition for conformality is the vanishing of the Cotton tensor; when $n\geq 4$, a necessary and sufficient condition for conformality is the vanishing of the Weyl tensor. In what follows we will assume that the corresponding conformality condition is always satisfied as the initial condition for the Ricci flow equations (\ref{muyrancio}). Then, with respect to the initial metric, $\mathbb{M}$ admits isothermal coordinates, that we continue to denote by $x^{i}$ and that we will use systematically in what follows. Therefore let the initial metric be
\begin{equation}
g_{ij}(t_0)={\rm e}^{-f}\delta_{ij},
\label{metrik}
\end{equation}
where $f$ is a smooth real function on $\mathbb{M}$, hereafter referred to as {\it conformal factor}\/.\footnote{Some authors use the term {\it conformal factor}\/ for the exponential ${\rm e}^{-f}$.} We will see presently that the Ricci flow of the metric (\ref{metrik}) leads to the metric
\begin{equation}
g_{ij}(t)={\rm e}^{-Et}g_{ij}(t_0),
\label{endlich}
\end{equation}
with $E$ a constant to be identified with the energy of the mechanical system under construction (see (\ref{getrennt}) below). For the Christoffel symbols corresponding to the metric (\ref{metrik}) we find
\begin{equation}
\Gamma_{ij}^m=\frac{1}{2}\left(\delta_{ij}\delta^{mh}\partial_hf-\delta^m_j\partial_{i}f-\delta_i^m\partial_jf\right),
\label{symbole}
\end{equation}
while the volume element equals
\begin{equation}
{\rm d}V={\rm e}^{-nf/2}{\rm d}^nx.
\label{skalar}
\end{equation}
A computation gives the Ricci tensor corresponding to the metric (\ref{metrik}),
\begin{equation}
R_{im}=\frac{1}{2}{\rm e}^{-f}\,\left(\frac{2-n}{2}\partial_jf\partial^jf+\partial_j\partial^jf\right)\delta_{im}+\frac{n-2}{2}\left(\partial_i\partial_mf+\frac{1}{2}\,\partial_if\partial_mf\right),
\label{ritchie}
\end{equation}
whose Ricci scalar $R=g^{im}R_{im}$ is
\begin{equation}
R=(n-1)\left(\frac{2-n}{4}\partial_jf\partial^jf+\partial_j\partial^jf\right).
\label{rico}
\end{equation}
Moreover, for any smooth function $\varphi$ we have
\begin{equation}
\nabla^2\varphi=\left(1-\frac{n}{2}\right)\partial_jf\partial^j\varphi+\partial_j\partial^j\varphi,
\qquad
\left(\nabla\varphi\right)^2=\left(\partial_j\varphi\right)\left(\partial^j\varphi\right).
\label{drei}
\end{equation}
In particular, setting $\varphi=f$ we arrive at
\begin{equation}
\partial_j\partial^jf=\nabla^2f+\left(\frac{n}{2}-1\right)(\nabla f )^2.
\label{kkonv}
\end{equation}
Using (\ref{kkonv}) in (\ref{ritchie}) and (\ref{rico}) we find the following equivalent expressions for the Ricci tensor $R_{im}$ and the Ricci scalar $R$:
\begin{equation}
R_{im}=\frac{1}{2}{\rm e}^{-f}\nabla^2f\delta_{im}+\frac{n-2}{2}\left(\partial_i\partial_m f+\frac{1}{2}\partial_if\partial_mf\right)
\label{masrichi}
\end{equation}
and
\begin{equation}
R=(n-1)\left[\frac{n-2}{4}(\nabla f)^2+\nabla^2f\right].
\label{ricox}
\end{equation}
In what follows we will find (\ref{masrichi}) and (\ref{ricox}) more convenient to use than their equivalents (\ref{ritchie}) and (\ref{rico}).

Compactness of $\mathbb{M}$ ensures the convergence of the integrals we will work with, without the need to impose further conditions on the integrands  (such as, {\it e.g.}, fast decay at infinity). The absence of a boundary ensures the possibility of integrating by parts without picking up boundary terms.

\section{Perelman's functional}\label{grisha}

We will first present some necessary background material \cite{PERELMAN, TOPPING}. Given a metric $g_{ij}$ and a real scalar field $\varphi$ on the manifold $\mathbb{M}$, Perelman's functional, denoted ${\cal F}[\varphi,g_{ij}]$, is defined as
\begin{equation}
{\cal F}[\varphi,g_{ij}]:=\int_{\mathbb{M}}{\rm e}^{-\varphi}\left[\left(\nabla \varphi\right)^2+ R(g_{ij})\right]{\rm d}V.
\label{epe}
\end{equation}
The gradient flow of ${\cal F}$ is given by the evolution equations
\begin{equation}
\frac{\partial g_{ij}}{\partial t}=-2\left(R_{ij}+\nabla_i\nabla_j\varphi\right),\qquad \frac{\partial\varphi}{\partial t}=-\nabla^2\varphi-R.
\label{rancio}
\end{equation}
Via a time--dependent diffeomorphism, the above are equivalent to
\begin{equation}
\frac{\partial g_{ij}}{\partial t}=-2R_{ij}, \qquad \frac{\partial \varphi}{\partial t}=-\nabla^2\varphi+\left(\nabla\varphi\right)^2-R.
\label{muyrancio}
\end{equation}
We will use (\ref{muyrancio}) rather than (\ref{rancio}). By  (\ref{skalar}) and (\ref{ricox}) we have, when the metric is conformal as in (\ref{metrik}),
\begin{equation}
{\cal F}[\varphi, f]:={\cal F}[\varphi,g_{ij}={\rm e}^{-f}\delta_{ij}]
\label{todorancio}
\end{equation}
$$
=\int_{\mathbb{M}}{\rm e}^{-\varphi-nf/2}\left\{(\nabla \varphi)^2+(n-1)\left[\frac{n-2}{4}(\nabla f)^2+\nabla^2f\right]\right\}{\rm d}^nx.
$$
In what follows we set $\varphi=f$, so the above functional simplifies to
\begin{equation}
{\cal F}[f]:={\cal F}[\varphi=f,f]=\int_{\mathbb{M}}{\rm e}^{-(1+\frac{n}{2})f}\,\left[a_n\left(\nabla f\right)^2+b_n\nabla^2f\right] {\rm d}^nx,
\label{epep}
\end{equation}
where the coefficients $a_n$ and $b_n$ are given by
\begin{equation}
a_n=1+\frac{(n-1)(n-2)}{4},\qquad b_n=n-1.
\label{ugly}
\end{equation}
Despite its nonstandard appearance, the coefficient $a_n$ of the kinetic term above is always positive as it should. After setting $\varphi=f$ we appear to have a contradiction, since we have two different flow equations in (\ref{muyrancio}) for just one field $f$. That there is in fact no contradiction can be seen as follows. In (\ref{muyrancio}) we have two different flow equations for two independent fields $f$ and $\varphi$. Equating the latter two fields implies that the two flow equations must reduce to just one. This can be achieved by substituting one of the two flow equations (\ref{muyrancio}) into the remaining one. Thus contracting the first of eqns.  (\ref{muyrancio}) with $g^{ij}$ we find
\begin{equation}
\frac{n}{2}\frac{\partial f}{\partial t}=R,
\label{calor}
\end{equation}
which, substituted into the second of (\ref{muyrancio}) leads to
\begin{equation}
\frac{\partial  f}{\partial t}+\frac{2}{2+n}\nabla^2  f-\frac{2}{2+n}\left(\nabla f\right)^2=0.
\label{sintildes}
\end{equation}
We will later on find it convenient to distinguish notationally between time--independent and time--dependent quantities. We thus rewrite our flow equation (\ref{sintildes}) as
\begin{equation}
\frac{\partial \tilde f}{\partial t}+\frac{2}{2+n}\nabla^2 \tilde f-\frac{2}{2+n}\left(\nabla \tilde f\right)^2=0,
\label{konvfluss}
\end{equation}
where a tilde on top of the conformal factor distinguishes it from the time--independent $f$ present in the functional (\ref{epep}).  One can verify that (\ref{konvfluss}) is the gradient flow equation of the functional ${\cal F}[\phi]$, where ${\cal F}$ is given as in (\ref{epep}) above, and $\phi$ is related to $f$ via a rescaling: $f=\frac{6-n}{4}\phi$.

\section{Proof of the theorem}\label{buch}

In what follows we regard the manifold $\mathbb{M}$ as the configuration space of a mechanical system, to be identified presently in terms of the conformal metric. We recall that, for a point particle of mass $m$ subject to a time--independent potential $U$, the Hamilton--Jacobi equation for the time--dependent action $\tilde S$ reads  \cite{ARNOLD}
\begin{equation}
\frac{\partial \tilde S}{\partial t}+\frac{1}{2m}\left(\nabla \tilde S\right)^2+U=0.
\label{hamjacbtrev}
\end{equation}
It is also well known that, separating the time variable as per
\begin{equation}
\tilde S=S-Et,
\label{trennung}
\end{equation}
with $S$ the so--called reduced action (also called Hamilton's characteristic function), one obtains
\begin{equation}
\frac{1}{2m}\left(\nabla S\right)^2+U=E.
\label{stillnight}
\end{equation}
Eqn. (\ref{trennung}) suggests separating variables in (\ref{konvfluss}) as per
\begin{equation}
\tilde f=f+Et,
\label{getrennt}
\end{equation}
where the sign of the time variable is reversed\footnote{This time reversal is imposed on us by the time--flow eqn. (\ref{konvfluss}), with respect to which time is reversed in the mechanical model. This is just a rewording of (part of) section 6.4 of ref. \cite{TOPPING}, where a corresponding heat flow is run {\it backwards}\/ in time.}
 with respect to (\ref{trennung}). Substituting (\ref{getrennt}) into  (\ref{konvfluss}) leads to
\begin{equation}
\frac{2}{2+n}\left(\nabla f\right)^2-\frac{2}{2+n}\nabla^2f=E
\label{masymas}
\end{equation}
which, upon using (\ref{ricox}), becomes
\begin{equation}
\frac{1}{2}\left(\nabla f\right)^2+\frac{2}{(2+n)(1-n)}R=E.
\label{novva}
\end{equation}
Comparing (\ref{novva}) with  (\ref{stillnight}) and  picking a value of the mass $m=1$, we conclude that the following identifications can be made:
\begin{equation}
S=f, \qquad U= \frac{2}{(2+n)(1-n)}R.
\label{geoint}
\end{equation}
So the potential $U$ is proportional to the scalar Ricci curvature of the configuration space  $\mathbb{M}$, while the reduced action $S$ equals the conformal factor $f$. This identifies a mechanical system in terms of the metric, such that the gradient Ricci flow of the latter equals the time--dependent Hamilton--Jacobi equation of the former. The theorem is proved.

\section{Proof of the corollary}\label{diskk}

We have so far considered the {\it classical}\/ mechanics associated with a given conformal factor, but one can immediately construct the corresponding {\it quantum}\/ mechanics, by means of the Schroedinger equation for the potential $U$. In fact the spectral problem for time--independent Schroedinger operators with the Ricci scalar as a potential function has been analysed in \cite{TOPPING}. So let us consider the time--dependent Schroedinger equation
\begin{equation}
{\rm i}\hbar\frac{\partial \psi}{\partial t}=-\frac{\hbar^2}{2m}\nabla^2\psi+U\psi,
\label{tdseq}
\end{equation}
which reduces to the time--dependent Hamilton--Jacobi equation (\ref{hamjacbtrev}) upon setting $\psi={\rm e}^{{\rm i}\tilde S/\hbar}$ and letting $\hbar\to 0$ \cite{ARNOLD}. We will hereafter use $\hbar=1$. Now (\ref{tdseq}) can be obtained as the extremal of the action functional
\begin{equation}
{\cal S}[\psi,\psi^*]:=\int_{\mathbb{M}}{\rm e}^{-nf/2}\left({\rm i}\psi^*\frac{\partial\psi}{\partial t}-\frac{1}{2m}\nabla\psi^*\nabla\psi-U\psi^*\psi\right){\rm d}^nx.
\label{actwelle}
\end{equation}
Substitute $\psi={\rm e}^{{\rm i}\tilde f}$ and $m=1$  above, use eqns. (\ref{ricox}) and (\ref{geoint}),  and finally consider the stationary case $\partial_t \tilde f=0$, where $\tilde f$ becomes $f$. Then (\ref{actwelle}) becomes
\begin{equation}
{\cal S}[f]:={\cal S}[\psi={\rm e}^{{\rm i}f}]=\frac{2}{n+2}\int_{\mathbb{M}}{\rm e}^{-nf/2}\left[-(\nabla f)^2+\nabla^2f\right]{\rm d}^nx.
\label{esseprima}
\end{equation}
Given the conformal factor $f$, define its rescaled $f_n$ as
$$
f_n:=\frac{n}{n+2}f.
$$
Then the functional (\ref{epep}), evaluated on $f_n$, can be expressed as the following integral containing the original conformal factor $f$:
\begin{equation}
{\cal F}[f_n]=\frac{n}{n+2}\int_{\mathbb{M}}{\rm e}^{-nf/2}\left[\frac{na_n}{n+2}\left(\nabla f\right)^2+b_n\nabla^2f\right]{\rm d}^nx.
\label{masefes}
\end{equation}
On the other hand, the Einstein--Hilbert gravitational action functional ${\cal G}$ on $\mathbb{M}$ is
\begin{equation}
{\cal G}[g_{ij}]:=\int_{\mathbb{M}}R(g_{ij})\,{\rm d}V,
\label{ggenerall}
\end{equation}
which, acting on the conformal metric (\ref{metrik}), becomes by (\ref{skalar}) and (\ref{ricox})
\begin{equation}
{\cal G}[f]:={\cal G}[g_{ij}(f)]=(n-1)\int_{\mathbb{M}}{\rm e}^{-nf/2}\left[\frac{n-2}{4}\left(\nabla f\right)^2+\nabla^2f\right]{\rm d}^nx.
\label{hilberteinstein}
\end{equation}
Now some algebra shows that
\begin{equation}
{\cal G}[f]={\cal F}[f_n]+{\cal S}[f]+{\cal C}[f],
\label{irreplaceble}
\end{equation}
where
\begin{equation}
{\cal C}[f]:=2c_n\int_{\mathbb{M}}{\rm e}^{-nf/2}\left[\frac{1}{2}\left(\nabla f\right)^2+d_n\nabla^2f\right]{\rm d}^nx,
\label{culo}
\end{equation}
is the action functional for a free scalar field $f$ coupled to a Coulomb charge placed at infinity \cite{DIFRANCESCO},  and the coefficients $c_n$, $d_n$ are given by
\begin{equation}
c_n=\frac{n^3-3n^2+n+6}{(n+2)^2}, \qquad d_n=\frac{(n-2)(n+2)}{n^3-3n^2+n+6}.
\label{east}
\end{equation}
The corollary is proved.

\section{Concluding remarks}

Above, the Einstein--Hilbert gravitational functional ${\cal G}[f]$, Schroedinger's action functional ${\cal S}[f]$ and the Coulomb functional ${\cal C}[f]$, as well as Perelman's entropy functional ${\cal F}[f]$, are all dimensionless. Introducing Newton's constant $G$, Planck's constant $\hbar$ and Boltzmann's constant $k$ restores their usual dimensions. Thus,  as announced, on a compact, conformally flat Riemannian configuration space without boundary, Einstein--Hilbert gravity arises from Schroedinger quantum mechanics, from Perelman's functional for the Ricci flow, and from the Coulomb functional. Since our starting point is ${\cal F}[f]$, the previous corollary is to be read in reverse: on a conformally flat manifold,  Perelman's functional contains Einstein--Hilbert gravity and Schroedinger quantum mechanics. 

The equality $f=S$ between the conformal factor $f$ and Hamilton's principal function $S$ trivially implies ${\rm e}^{-f}={\rm e}^{-S}$, but there are some nontrivialities concealed behind. In Feynman's formulation of quantum mechanics, Feynman's kernel in Euclidean space is given precisely by ${\rm e}^{-S}$. In other words, the conformal metric ${\rm e}^{-f}\delta_{ij}$ and its Ricci flow know about Feynman's quantum--mechanical kernel. This appears to suggest some deeper connection between the Ricci flow, gravity, quantum mechanics, and an approach to the latter two known as {\it emergent}\/ quantum mechanics. Interesting links between geometry and  quantum mechanics have been analysed in refs. \cite{MATONE, KOCH1, KOCH2, KOCH3}; the role of conformal symmetry and Ricci flow has been extensively studied in refs. \cite{CARROLL0, CARROLL1, CARROLL5, CARROLL2}; the closely related emergent quantum mechanics has been dealt with in refs. \cite{THOOFT1, THOOFT2, ELZE1, ELZE5}.

To round up our discussion let us point out some analogies between our approach and those just quoted. As stressed above, our starting point is Perelman's functional ${\cal F}[f]$, which is close, though not exactly identical, to the action functionals considered in \cite{KOCH1, KOCH2, KOCH3}. In particular, we have only considered the phase of the wavefunction,  reserving the amplitude for future study \cite{NOI}. Noteworthy features that our approach shares in common with that of \cite{KOCH1, KOCH2, KOCH3} are the natural appearance of the Ricci scalar curvature as the potential function, and the key role played by conformal symmetry. The issue of emergent quantum mechanics  in connection with the Ricci flow has been touched upon in \cite{ISIDRO, DICE}; this is a promising research line that deserves further attention.\\
\newpage

\noindent
{\bf Acknowledgements} Technical discussions with H.-T. Elze and B. Koch are gratefully acknowledged. J.M.I. is pleased to thank Max--Planck--Institut f\"ur Gravitationsphysik, Albert--Einstein--Institut (Potsdam, Germany) for hospitality extended over a long period of time.---{\it  Irrtum verl\"asst uns nie, doch ziehet ein h\"oher Bed\"urfnis immer den strebenden Geist leise zur Wahrheit hinan.}  Goethe.


\begin{thebibliography}{99}


\bibitem{SERGIO}
S. Abraham, P. Fern\'andez de C\'ordoba, J.M. Isidro and J.L.G. Santander, {\it The Ricci flow on Riemann surfaces}, {\tt arXiv:0810.2236 [hep-th]}.

\bibitem{ARNOLD}
V. Arnold, {\it Mathematical Methods of Classical Mechanics}, Springer, Berlin (1991).

\bibitem{MATONE}
G. Bertoldi, A. Faraggi and M. Matone, {\it Equivalence principle, higher dimensional Mobius group and the hidden antisymmetric tensor of quantum mechanics},
Class. Quant. Grav. {\bf 17} (2000) 3965, {\tt arXiv:hep-th/9909201}.

\bibitem{CARROLL0}
R. Carroll, {\it Fisher, K\"ahler, Weyl, and the quantum potential}, {\tt arXiv: quant-ph/0406210}.

\bibitem{CARROLL1}
R. Carroll, {\it Some remarks on Ricci flow and the quantum potential}, {\tt arXiv:math-ph/0703065}.

\bibitem{CARROLL5}
R. Carroll, {\it Remarks on Weyl geometry and quantum mechanics}, {\tt arXiv:0705.3921 [gr-qc]}.

\bibitem{CARROLL2}
R. Carroll, {\it Ricci flow and quantum theory}, {\tt arXiv:0710.4351 [math-ph]}.

\bibitem{CARROLL6}
R. Carroll, {\it Remarks on the Friedman equations}, {\tt arXiv:0712.3251 [math-ph]}.

\bibitem{DHOKER}
E. D'Hoker, {\it String Theory}, in {\it Quantum Fields and Strings: A Course for Mathematicians}, vol. 2, American Mathematical Society, Providence (1999).

\bibitem{DIFRANCESCO}
P. di Francesco, P. Mathieu and D. S\'en\' echal, {\it Conformal Field Theory}, Springer, Berlin (1997).

\bibitem{ELZE1}
H.-T. Elze, {\it Note on the existence theorem in ``Emergent quantum mechanics and emergent symmetries"}, {\tt arXiv:0710.2765 [quant-ph]}.

\bibitem{ELZE5}
H.-T. Elze, {\it The attractor and the quantum states}, {\tt arXiv:0806.3408 [quant-ph]}.

\bibitem{GOLDBERG}
S. Goldberg, {\it Curvature and Homology}, Dover, New York (1982).

\bibitem{ISIDRO}
J.M. Isidro, J.L.G. Santander and P. Fern\'andez de C\'ordoba, {\it Ricci flow, quantum mechanics and gravity},
{\tt arXiv:0808.2351 [hep-th]}.

\bibitem{RICCI}
J.M. Isidro, J.L.G. Santander and P. Fern\'andez de C\'ordoba, {\it A note on the quantum--mechanical Ricci flow},
{\tt arXiv:0808.2717 [hep-th]}.

\bibitem{DICE}
J.M. Isidro, J.L.G. Santander and P. Fern\'andez de C\'ordoba, {\it On the Ricci flow and emergent quantum mechanics}, {\tt arXiv:0902.0143 [hep-th]}.

\bibitem{NOI}
J.M. Isidro, J.L.G. Santander and P. Fern\'andez de C\'ordoba, in preparation.

\bibitem{KOCH1}
B. Koch, {\it Geometrical interpretation of the quantum Klein--Gordon equation}, {\tt arXiv: 0801.4635 [quant-ph]}.

\bibitem{KOCH2}
B. Koch, {\it Relativistic Bohmian mechanics from scalar gravity}, {\tt  arXiv:0810.2786 [hep-th]}.

\bibitem{KOCH3}
B. Koch, {\it A geometrical dual to relativistic Bohmian mechanics - the multi particle case}, {\tt arXiv:0901.4106 [gr-qc]}.

\bibitem{PERELMAN}
G. Perelman, {\it The entropy formula for the Ricci flow and its geometric applications}, {\tt arXiv:math/0211159 [math.DG]}.

\bibitem{THOOFT1}
G. 't Hooft, {\it The mathematical basis for deterministic quantum mechanics}, {\tt arXiv:quant-ph/0604008}.

\bibitem{THOOFT2}
G. 't Hooft, {\it Emergent quantum mechanics and emergent symmetries}, {\tt arXiv:0707.4568 [hep-th]}.

\bibitem{TOPPING}
P. Topping, {\it Lectures on the Ricci Flow}, London Mathematical Society Lecture Notes Series {\bf 325}, Cambridge University Press (2006).

\bibitem{WEYL}
H. Weyl, {\it Space, Time, Matter}, Dover, New York (1952).




\end{thebibliography}
\end{document}